\documentclass[10pt,a4paper]{article}
\usepackage[a4paper, left=2cm, right=2cm, top=1cm]{geometry}

\usepackage[utf8]{inputenc}
\usepackage{amsmath, amsthm, amssymb, amsfonts}     
\usepackage{mathrsfs}                               
\usepackage{accents}                                
\usepackage{natbib}                                 
\usepackage{hyperref}                               
\usepackage{doi}                                    
\usepackage{authblk}                                
\usepackage{siunitx}                                
\usepackage{booktabs}                               
\usepackage{ulem}                                   
\usepackage{nomencl}                                
\makenomenclature                                   
\usepackage[linesnumbered,ruled,vlined]{algorithm2e}
\SetCommentSty{mycommfont}
\usepackage{subfigure}

\usepackage{bm}                     

\usepackage{graphicx}
\usepackage{tikz}
\usepackage{pgfplots}

\usetikzlibrary{decorations.pathmorphing,arrows,intersections,arrows.meta,angles,quotes}

\pgfplotsset{
	kurze Legende/.style={
		legend image code/.code={
			\draw[##1,mark repeat=2,line width=0.6pt]
			plot coordinates {
				(0cm,0cm)
				(0.3cm,0cm)
			};
		}
	}
}

\pgfplotsset{
	compat = newest,
	scale only axis, 
	max space between ticks = 50pt,
	ticklabel style = {font=\footnotesize},
	legend style =  {font=\footnotesize},
	grid = major,
	grid style = {dotted},
	legend columns=1, 
	xtick pos=left,
	ytick pos=left
}

\newcommand{\tinyPicture}{
	\pgfplotsset{  
		width=0.3\textwidth,
		height=0.3\textwidth,
		ylabel style={text width=0.2\textwidth,align=center}
	}
}

\pgfplotsset{select coords between index/.style 2 args={
		x filter/.code={
			\ifnum\coordindex<#1\fi
			\ifnum\coordindex>#2\fi
		}
}}

\definecolor{color1}{HTML}{0060AD} 
\definecolor{color2}{HTML}{FF4500} 
\definecolor{color3}{HTML}{FFA500} 
\definecolor{color4}{HTML}{006400} 
\definecolor{color5}{HTML}{9400D3} 
\definecolor{color6}{HTML}{800000} 
\definecolor{color7}{HTML}{000000} 
\definecolor{color8}{HTML}{0000FF} 
\definecolor{color9}{HTML}{FF0000} 
\definecolor{mycolor_blue}{RGB}{66,124,161}
\definecolor{mycolor_grey}{RGB}{198,198,198} 

\tikzstyle{line1} = [color=color7,semithick] 
\tikzstyle{line2} = [color=color2,densely dotted,semithick]
\tikzstyle{line3} = [color=color1,densely dashed,semithick]
\tikzstyle{line4} = [color=color5,dash dot,semithick]
\tikzstyle{line5} = [color=color4,dash dot dot,semithick]
\tikzstyle{line6} = [color=color6,semithick]

\tikzstyle{mark1} = [color=color7,mark=x,mark size=2pt,mark options=solid,semithick] 
\tikzstyle{mark2} = [color=color2,mark=o,mark size=2pt,mark options=solid,semithick]
\tikzstyle{mark3} = [color=color1,mark=*,mark size=2pt,mark options=solid,semithick]
\tikzstyle{mark4} = [color=color5,mark=triangle,mark size=2pt,mark options=solid,semithick]
\tikzstyle{mark5} = [color=color4,mark=square,mark size=2pt,mark options=solid,semithick]
\tikzstyle{mark6} = [color=color7,mark=o,mark size=2pt,mark options=solid,semithick]
\tikzstyle{mark7} = [color=color7,mark=*,mark size=2pt,mark options=solid,semithick]
\tikzstyle{mark8} = [color=color7,mark=triangle,mark size=2pt,mark options=solid,semithick]

\usetikzlibrary{external}
\title{Identification of Beneficial and Detrimental Structure Locations Around Flettner Rotors Using Topology-Optimization-Inspired Sensitivity Fields}
\author[]{Niklas K\"uhl\thanks{kuehl@hsva.de}}

\affil[]{Hamburg Ship Model Basin, Bramfelder Strasse 164, D-22305 Hamburg, Germany}

\begin{document}

\providetoggle{tikzExternal}
\settoggle{tikzExternal}{true}
\settoggle{tikzExternal}{false}

\maketitle

\begin{abstract}
Flettner rotors are highly sensitive to their surrounding flow field and may be significantly affected by nearby ship structures, deck cargo, and superstructures. Assessing the aerodynamic influence of such structures during early design stages remains challenging, particularly when a large number of potential arrangements must be considered.

This paper presents a topology-optimization-inspired numerical sensitivity-analysis approach for identifying beneficial and detrimental locations of additional structures around Flettner rotors. The numerical method is based on a virtual porosity formulation and evaluates the corresponding sensitivity field using a continuous adjoint framework. In contrast to classical topology optimization, the porosity field is not treated as a design variable and no optimization loop is performed. Instead, the resulting sensitivity field is interpreted as a design-support tool that indicates regions where the introduction of material is expected to improve or deteriorate a selected aerodynamic objective.

The approach is demonstrated for a full-scale Flettner rotor operating at a diameter-based Reynolds number of $\mathrm{Re_D}=2\times10^6$ and a spinning ratio of $k=3$. Sensitivity fields are evaluated for drag, lift, and a combined objective. Their predictive capability is assessed by positioning container stacks at locations identified as beneficial or detrimental by the sensitivity analysis and subsequently re-evaluating the aerodynamic performance of the modified configurations.

The obtained results show that the predicted sensitivity regions correctly indicate the qualitative influence of nearby structures on rotor performance. The proposed methodology therefore provides an efficient screening and design-support tool for assessing structure-placement effects around Flettner rotors and may support the aerodynamic integration of wind-assisted propulsion systems during early design stages.
\end{abstract}

\begin{flushleft}
\end{flushleft}

\section{Introduction}
\label{sec:introduction}
Flettner rotors are increasingly considered a promising technology for wind-assisted ship propulsion and have attracted considerable attention in the context of reducing fuel consumption and greenhouse gas emissions in maritime transportation. Their aerodynamic performance strongly depends on the local flow conditions around the rotor and may therefore be significantly affected by nearby ship structures such as deckhouses, cargo stacks, funnels, cranes, or neighboring rotors, cf. \cite{albers2026case, bordogna2020effects, hoffmann2026numerical}. While such interactions can deteriorate rotor performance, they may also lead to beneficial flow modifications. Identifying favorable and unfavorable arrangements therefore represents an important task during the aerodynamic integration of Flettner rotors into ship designs.

Traditionally, the aerodynamic influence of surrounding structures is assessed through dedicated Computational Fluid Dynamics (CFD) studies in which individual design variants are investigated and compared, e.g., \cite{saydam2018evaluation, el2023experimental}. However, the large number of possible arrangements quickly renders exhaustive parameter studies impractical during early design stages. Consequently, there is a need for methods that provide guidance on potentially beneficial and detrimental structure locations without requiring a comprehensive design-space exploration.

In a broader context, topology optimization methods have demonstrated a remarkable capability to identify favorable material distributions within fluid domains, cf. \cite{papoutsis2016continuous, bendsoe2013topology}. In particular, porous-medium-based formulations combined with adjoint sensitivity analysis have been successfully applied to a wide range of fluid-mechanical optimization problems, including creeping flows, ducted flow applications, and aerodynamic configurations (\cite{borrvall2003topology, guest2006topology, gersborg2005topology, othmer2008continuous, gerdes2014efficient, othmer2017aerodynamic, vrionis2021topology}). Such approaches typically treat the porosity distribution as a design variable and employ gradient-based optimization algorithms to iteratively evolve the domain topology towards improved performance.

The present work follows a different philosophy. Rather than performing a topology optimization, a virtual porosity parameter is introduced solely for the purpose of evaluating local sensitivity information. The corresponding sensitivity field is obtained through a continuous adjoint formulation (\cite{giles1997adjoint, giles2000introduction, othmer2008continuous}) and interpreted as an indicator of beneficial and detrimental regions for the placement of additional structures. The porosity field is therefore not used as a design variable and no optimization loop is performed. Instead, the resulting sensitivity information serves as a design-support and screening tool for assessing the qualitative influence of material placement on selected aerodynamic objectives.

To demonstrate the practical applicability of the proposed approach, a full-scale Flettner rotor operating under atmospheric wind conditions is investigated. Sensitivity fields are evaluated for drag, lift, and a combined objective function. Their predictive capability is subsequently assessed by positioning container stacks at locations identified as beneficial or detrimental by the sensitivity analysis and re-evaluating the aerodynamic performance of the modified configurations.

The remainder of the manuscript is structured as follows. Section \ref{sec:mathematical_model} summarizes the governing equations and the corresponding sensitivity formulation. Subsequently, Section \ref{sec:application} applies the proposed methodology to a full-scale Flettner rotor configuration and assesses its predictive capability through several container-placement studies. Finally, conclusions and an outlook on future work are provided in Section \ref{sec:conclusion_outlook}.

\section{Mathematical Model}
\label{sec:mathematical_model}
The proposed methodology is based on the formal introduction of a Darcy-type penalty term into the incompressible Navier--Stokes equations. While the penalty parameter remains inactive throughout all primal flow simulations, it serves as an auxiliary quantity for deriving local sensitivity information regarding the introduction of additional material into the flow domain. The governing equations for the incompressible flow field read
\begin{align}
-\frac{\partial v_k}{\partial x_k} &= 0 \label{equ:mass_balance} ,  \\
\frac{\partial \rho v_i}{\partial t} + \frac{\partial v_k \rho v_i}{\partial x_k} + \frac{\partial}{\partial x_k} \left[ p^\mathrm{eff} \delta_{ik} - 2 \mu^\mathrm{eff} S_{ik} \right] + \alpha \left( v_i - v_i^\mathrm{tar} \right) &= 0 ,  \label{equ:momentum_balance}
\end{align}
where $v_i=v_i(x_i,t)$ and $\rho$ denote the fluid velocity and density, respectively. Therein, $x_i$ and $t$ represent the spatial and temporal coordinates. Furthermore, $S_{ik}=1/2 (\partial v_i/\partial x_k + \partial v_k/\partial x_i)$ represents the strain-rate tensor, while $p^\mathrm{eff}=p+p^t$ and $\mu^\mathrm{eff}=\mu+\mu^t$ denote the effective pressure and viscosity, respectively, composed of molecular and turbulence-model contributions. Turbulent quantities are approximated using a hybrid averaged/filtered turbulence modeling approach based on the Improved Delayed Detached-Eddy Simulation (IDDES) model of \cite{gritskevich2012development}. Further details regarding the employed turbulence modeling are provided in the appendices of \cite{kuhl2025incremental, kuhl2025differential}.

The parameter $\alpha(x_i) \geq 0$ denotes a Darcy-type penalty coefficient that drives the flow towards a prescribed target velocity $v_i^\mathrm{tar}$. In classical porous-medium formulations, the target velocity usually vanishes, i.e. $v_i^\mathrm{tar}=0$, such that the additional source term acts as a momentum sink. In the present work, however, the penalty parameter is not employed as a physical model quantity and remains zero throughout all flow simulations. Instead, it is introduced solely for the purpose of evaluating sensitivities with respect to a virtual material distribution. Alternatives based on Forchheimer-type formulations are also conceivable but are not considered in the present study.

The Flettner rotor is considered in an idealized numerical wind-tunnel configuration employing the following boundary conditions along the rotor surface ($\Gamma^\text{wall}$), the lower symmetry plane ($\Gamma^\text{symm}$), the horizontal far-field boundaries ($\Gamma^\text{hor}$), and the upper far-field boundary ($\Gamma^\text{top}$):
\begin{alignat}{3}
\frac{\partial p}{\partial n} &= 0, &&v_i = V_i^\mathrm{wall} \qquad &&\text{on} \qquad \Gamma^\text{wall} ,  \\
\frac{\partial p}{\partial n} &= 0, && v_i = V_i^\mathrm{hor} \qquad &&\text{on} \qquad \Gamma^\text{hor} ,  \\
\frac{\partial p}{\partial n} &= 0, &&\frac{\partial v_i}{\partial n} = 0 \qquad &&\text{on} \qquad \Gamma^\text{symm} ,  \\
p &= P, &&\frac{\partial v_i}{\partial n} = 0 \qquad &&\text{on} \qquad \Gamma^\text{top} , .
\end{alignat}
Therein, $P$ denotes a reference pressure and $V_i^\mathrm{hor}=V_\mathrm{ref}(x_3/x_{3,\mathrm{ref}})^m\delta_{i1}$ represents an atmospheric wind profile pointing in the $x_1$ direction and varying with the vertical coordinate $x_3$. The circumferential rotor velocity is prescribed as $V_i^\mathrm{wall}=(\pi D f)t_i$, where $D$, $f$, and $t_i$ denote the rotor diameter, rotational frequency, and circumferential unit vector, respectively. The initial conditions correspond to the atmospheric inflow profile extended throughout the computational domain, i.e.,
\begin{align}
p(x_i,t=0)=P
\qquad\text{and}\qquad
v_i(x_i,t=0)=V_i^\mathrm{hor} , .
\end{align}

The present study aims to identify beneficial and detrimental regions for material placement in the vicinity of a Flettner rotor by evaluating topological sensitivities of statistically converged flow fields. Since the underlying hybrid RANS/LES simulations are inherently unsteady, all forward calculations are carried out until statistical convergence of the time-averaged quantities is achieved. This motivates the following force-based objective functional
\begin{align}
J = \frac{1}{t^\mathrm{end} - t^\mathrm{start}} \int_{t^\mathrm{start}}^{t^\mathrm{end}} \int \left[ p^\mathrm{eff} \delta_{ik} - 2 \mu^\mathrm{eff} S_{ik} \right] n_k r_i \mathrm{d} \Gamma^\mathrm{O} \mathrm{d} t
\quad
\to
\quad
J = \int \left[ \bar{p}^\mathrm{eff} \delta_{ik} - 2 \bar{\mu}^\mathrm{eff} \bar{S}_{ik} \right] n_k r_i \mathrm{d} \Gamma^\mathrm{O}
\label{equ:objective}
\end{align}
where overlined quantities denote temporal averages and are therefore no longer functions of time, e.g., $\bar{v}_i=\bar{v}_i(x_i)$ and $\bar{p}=\bar{p}(x_i)$. Furthermore, $n_k$ denotes the local surface-normal vector and $r_i$ specifies the force direction under consideration, such as drag or lift. The objective functional is evaluated exclusively on a subset of the domain boundary, i.e., $\Gamma^\mathrm{O}\subset\Gamma^\mathrm{wall}\subset\Gamma$, which corresponds to the Flettner rotor surface.

In accordance with the time-averaged objective functional, the corresponding Lagrangian is constructed by augmenting the objective with the time-averaged governing equations (cf. Eqns. \eqref{equ:mass_balance} and \eqref{equ:momentum_balance}), i.e.,
\begin{align}
    L = J + \int \hat{v}_i \left[  \frac{\partial}{\partial x_k} \left[ \bar{v}_k \rho \bar{v}_i + \bar{p}^\mathrm{eff} \delta_{ik} - 2 \bar{\mu}^\mathrm{eff} \bar{S}_{ik} \right] + \alpha \left( \bar{v}_i - v_i^\mathrm{tar} \right) \right] - \hat{p} \frac{\partial \bar{v}_k}{\partial x_k} \mathrm{d} \Omega \, ,
\end{align}
where $\hat{v}_i$ and $\hat{p}$ denote the adjoint velocity and pressure fields, respectively. Their physical dimensions follow from the objective functional and read $[\hat{v}_i] = \SI{}{[J]/N}$ and $[\hat{p}] = \SI{}{[J]/(m^3/s)}$, implying a dimensionless adjoint velocity for the force objectives considered in this study. Owing to the time-averaged formulation, temporal momentum transport is omitted and the convective transport is incorporated into the divergence operator of the effective flow quantities. A variation of the extended cost functional reads
\begin{align}
    L^\prime = J^\prime + \int \hat{v}_i \left[  \frac{\partial}{\partial x_k} \left[ v_k^\prime \rho v_i + v_k \rho v_i^\prime + {p^\mathrm{eff}}^\prime \delta_{ik} - 2 \mu^\mathrm{eff} S_{ik}^\prime \right] + \alpha^\prime \left( v_i - v_i^\mathrm{tar} \right) + \alpha v_i^\prime \right] - \hat{p} \frac{\partial v_k^\prime}{\partial x_k} \mathrm{d} \Omega \, ,
\end{align}
where variations of the fluid density and effective turbulent viscosity have been neglected. The latter corresponds to the commonly employed frozen-turbulence assumption, which represents an approximation, particularly at the high Reynolds numbers considered in the present study (\cite{lohner2003adjoint, othmer2008continuous, stuck2013adjoint, kroger2018adjoint}). Nevertheless, the subsequent validation studies demonstrate that meaningful sensitivity information can still be obtained.
An isolation of all varied quantities can be developed to
\begin{align}
    L^\prime 
    &= \int {\bar{p}^\mathrm{eff}}^\prime n_k r_k - 2 \bar{\mu}^\mathrm{eff} \bar{S}_{ik}^\prime n_k r_i \mathrm{d} \Gamma^\mathrm{O} \label{equ:varied_objective} \\
    &+ \int \hat{v}_i \left[ v_k^\prime \rho v_i + v_k \rho v_i^\prime + {\bar{p}^\mathrm{eff}}^\prime \delta_{ik} - \bar{S}_{ik}^\prime 2 \bar{\mu}^\mathrm{eff} \right] n_k  - \hat{p} v_k^\prime n_k + \bar{\mu}^\mathrm{eff} \frac{\partial \hat{v}_i}{\partial x_k} \left( v_i^\prime n_k + v_k^\prime n_i \right) \mathrm{d} \Gamma \label{equ:adjoint_boundary_integrals} \\
    &+ \int v_i^\prime \left[ - \rho v_k \frac{\partial \hat{v}_k}{\partial x_i}- \rho v_k \frac{\partial \hat{v}_i}{\partial x_k} + \frac{\partial}{\partial x_k}\left[ \hat{p} \delta_{ik} - \bar{\mu}^\mathrm{eff} 2 \hat{S}_{ik} \right]  + \alpha \hat{v}_i \right] - {\bar{p}^\mathrm{eff}}^\prime  \frac{\partial \hat{v}_k}{\partial x_k} \label{equ:adjoint_field_integrals} \\
    &\qquad + \alpha^\prime \hat{v}_i \left( v_i - v_i^\mathrm{tar} \right) \mathrm{d} \Omega \, .
\end{align}
Requiring the first variation of the Lagrangian to vanish, i.e. $L^\prime = 0$, yields the following adjoint system of equations, which eliminates the volume integral in Eqn. \eqref{equ:adjoint_field_integrals}:
\begin{align}
    -\frac{\partial \hat{v}_k}{\partial x_k} &= 0 \label{equ:adjoint_mass_balance} \, , \\
    - \rho \bar{v_k} \frac{\partial \hat{v}_k}{\partial x_i}- \rho \bar{v_k} \frac{\partial \hat{v}_i}{\partial x_k} + \frac{\partial}{\partial x_k}\left[ \hat{p} \delta_{ik} - \bar{\mu}^\mathrm{eff} 2 \hat{S}_{ik} \right]  + \alpha \hat{v}_i &= 0 \, . \label{equ:adjoint_momentum_balance}
\end{align}
The required adjoint boundary conditions follow from the remaining boundary contributions in Eqns. \eqref{equ:varied_objective}--\eqref{equ:adjoint_boundary_integrals}, i.e.
\begin{alignat}{3}
    \frac{\partial \hat{p}}{\partial n} &= 0, &&\hat{v}_i = -r_i \qquad &&\text{on} \qquad \Gamma^\text{wall} \, , \label{equ:adwall_bc} \\
    \frac{\partial \hat{p}}{\partial n} &= 0, &&\hat{v}_i = 0 \qquad &&\text{on} \qquad \Gamma^\text{hor} \, , \\
    \frac{\partial \hat{p}}{\partial n} &= 0, &&\frac{\partial \hat{v}_i}{\partial n} = 0 \qquad &&\text{on} \qquad \Gamma^\text{symm} \\
    \hat{p} &= 0, &&\frac{\partial \hat{v}_i}{\partial n} = 0 \qquad &&\text{on} \qquad \Gamma^\text{top} \, .
\end{alignat}
Further details regarding continuous adjoint formulations can be found, for example, in \cite{stuck2012adjoint, kroger2016numerical}. The remaining terms of the varied Lagrangian provide access to the local topological sensitivity, i.e.,
\begin{align}
    L^\prime = \int \alpha^\prime \hat{v}_i \left( v_i - v_i^\mathrm{tar} \right) \mathrm{d} \Omega
    \qquad \qquad \overset{v_i^\mathrm{tar} = 0}{\to} \qquad \qquad
    s = \hat{v}_i v_i \, , \label{equ:topo_derivative}
\end{align}
which corresponds to the inner product of the primal and adjoint velocity vectors, cf. \cite{gerdes2018fluid}.

The sensitivity expression quantifies the influence of a local porosity perturbation on the objective functional. In classical topology optimization, this information is typically employed within an iterative optimization framework, where the porosity field is repeatedly updated and the flow solution re-evaluated until a suitable material distribution is obtained (\cite{gerdes2014efficient, vrionis2021topology}). Such procedures usually require additional constraints and regularization measures to ensure physically meaningful designs, for example, non-negative porosity values or minimum feature sizes (\cite{de2015stress, alexandersen2015topology}).

The present work follows a different philosophy. Rather than performing an actual topology optimization, only the sign of the sensitivity field is evaluated and interpreted as an indicator of beneficial and detrimental regions for material placement. The resulting sensitivity maps are therefore used as a screening and design-support tool that provides qualitative guidance during early design stages. In the context of the present study, this information is employed to assess suitable locations for container stacks in the vicinity of a Flettner rotor.

\section{Application}
\label{sec:application}

The proposed methodology is demonstrated for the flow around a Flettner rotor equipped with two end plates. The practical design task considered in the following refers to the placement of six Twenty-foot Equivalent Unit containers (TEU20) in the vicinity of the rotor. This represents a typical problem encountered in ship design, as Flettner rotors almost always interact with surrounding deck cargo and superstructures.

The governing equations are approximated using the segregated primal/adjoint flow solver FreSCo$^+$ (\cite{rung2009challenges, stuck2012adjoint, schubert2019analysis, manzke2018development}). The code employs an implicit second-order finite-volume formulation on unstructured polyhedral meshes and utilizes the SIMPLE algorithm for pressure--velocity coupling (\cite{rhie1983numerical, yakubov2015experience, kuhl2022discrete}). 

Diffusive fluxes are discretized using a second-order Central Difference Scheme (CDS), while convective fluxes follow a blended CDS/QUICK formulation employing 80\% CDS and 20\% QUICK. Near-wall treatment relies on the universal wall function of \cite{rung2001universal, gritskevich2017comprehensive} with a target wall resolution of $y^+ \approx 50$. Time integration is performed using a second-order Implicit Three-Time-Level (ITTL) scheme.

The employed primal and adjoint formulations have been extensively validated in previous studies. Representative validations for maritime aerodynamic applications can be found in \cite{kroger2016numerical, angerbauer2020hybrid, kroger2018adjoint}. Further implementation details are provided in the cited references.

\subsection{Reference Rotor Configuration}
\label{subsec:stand_alone}
An isolated full-scale Flettner rotor is considered at a diameter-based Reynolds number of $\mathrm{Re_D}=\rho V D/\mu=2\times10^6$ and a height-based Reynolds number of $\mathrm{Re_H}=\rho V H/\mu=1.2\times10^7$. The rotor operates at a spinning ratio of $k=\pi D n/V=3$, where $D$ and $H$ denote the rotor diameter and height, respectively, and $V$ refers to the reference wind speed evaluated at the reference height $x_{3,\mathrm{ref}}/H=10/18$. The rotor is located at $x_i=[0,0,0]^\mathrm{T}$, is exposed to an incoming flow directed along the negative $x_1$ axis, and rotates about the $x_3$ axis. To minimize blockage effects, all far-field boundaries are positioned 100 rotor diameters away from the rotor. The horizontal boundaries prescribe the atmospheric inflow profile introduced previously using a Hellmann exponent of $m=0.85$, while the upper and lower boundaries employ pressure and symmetry conditions, respectively.

\begin{figure}[!htb]
\centering
\subfigure[]{
\includegraphics[width=0.475\textwidth]{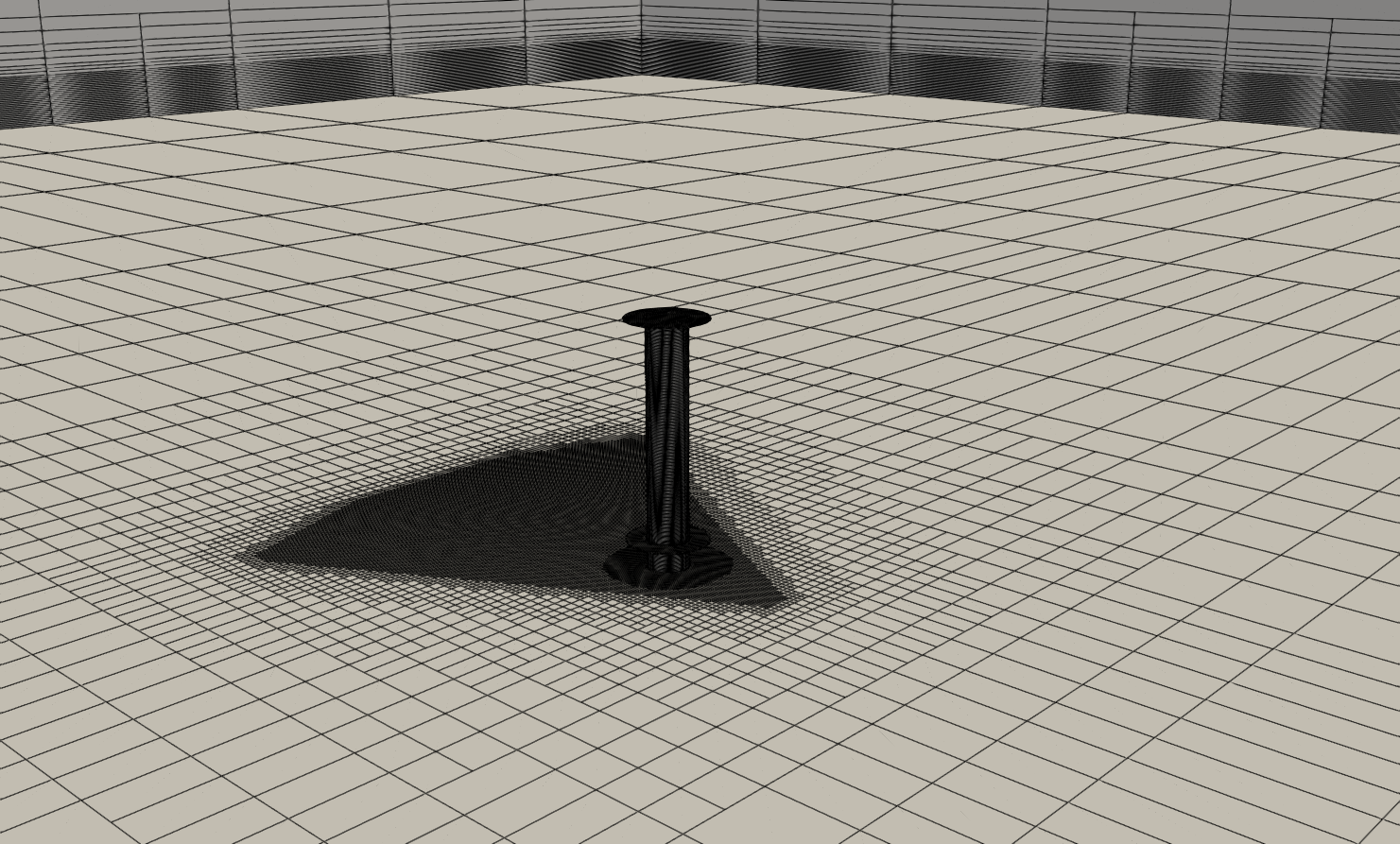}
}
\subfigure[]{
\includegraphics[width=0.425\textwidth]{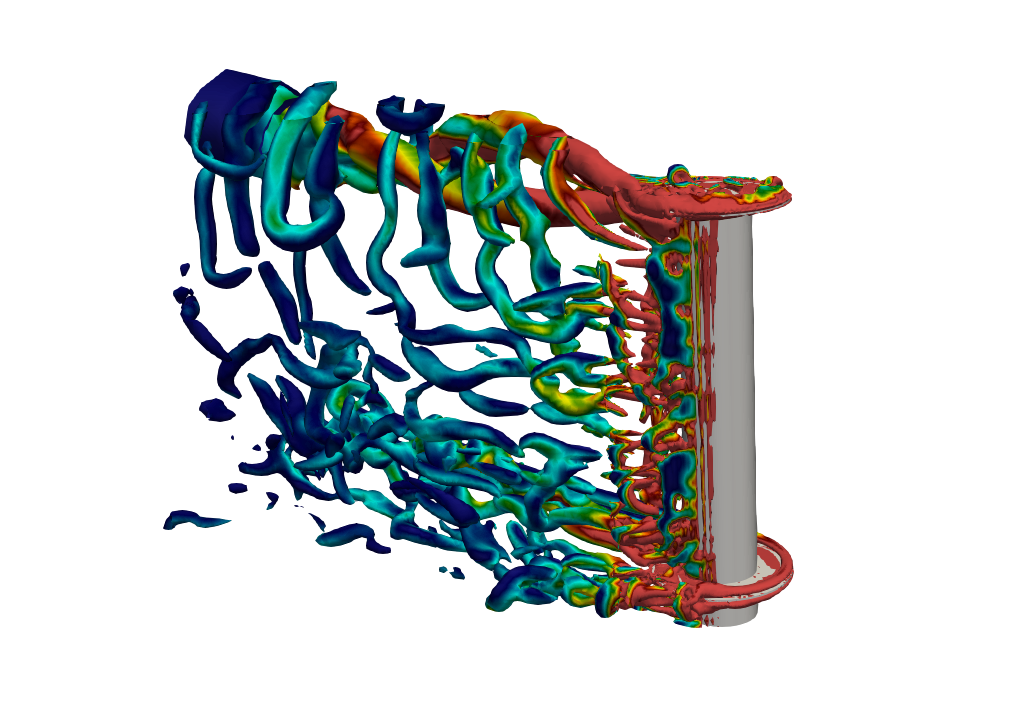}
}
\caption{(a) Near-field computational grid and (b) instantaneous vortex structures visualized by the $Q$-criterion and colored by vorticity magnitude for the reference rotor configuration at $\mathrm{Re_D}=2\times10^6$ and $k=3$.}
\label{fig:single_rotor_DES}
\end{figure}

An impression of the computational setup is provided in Fig.~\ref{fig:single_rotor_DES}(a), showing the unstructured near-field grid together with the wake and atmospheric-boundary-layer refinements. The numerical mesh consists of approximately 3.8 million control volumes. The simulations are performed for 200 equivalent rotor-flow passages while maintaining a Courant number below unity, resulting in approximately $\SI{20000}{}$ uniform time steps. Time averaging starts after 100 equivalent passages. Figure~\ref{fig:single_rotor_DES}(b) provides an instantaneous visualization of the resolved vortex structures using the $Q$-criterion ($Q=100$), colored by the local vorticity magnitude. Besides the smaller wake structures, the pronounced rotor tip vortex is clearly visible.

\begin{figure}[!ht]
\centering
\subfigure[]{
\iftoggle{tikzExternal}{
\input{./tikz/des_avg_impression.tikz}}{
\includegraphics{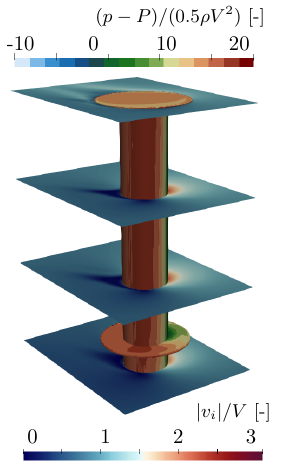}
}
}
\subfigure[]{
\iftoggle{tikzExternal}{
\input{./tikz/des_credibility.tikz}}{
\includegraphics{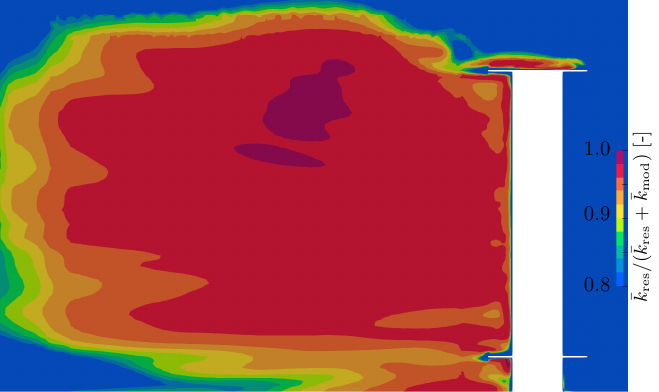}
}
}
\caption{(a) Normalized time-averaged velocity magnitude $|v_i|/V$ on four vertical sectional planes together with the pressure coefficient distribution along the rotor surface and (b) ratio of resolved ($k_\mathrm{res}$) to total ($k_\mathrm{res}+k_\mathrm{mod}$) turbulent kinetic energy in the rotor wake.}
\label{fig:des_credibility}
\end{figure}

Figure~\ref{fig:des_credibility}(a) provides an impression of the statistically converged flow field by showing the normalized time-averaged velocity magnitude $|v_i|/V$ on four vertical sectional planes at $x_3/H=[0,1/3,2/3,1]$ together with the corresponding pressure coefficient distribution $c_p=(p-P)/(0.5\rho V^2)$ along the rotor surface.
The credibility of the hybrid RANS/LES prediction is assessed in Fig.~\ref{fig:des_credibility}(b) by evaluating the ratio of resolved to total Turbulent Kinetic Energy (TKE) in the rotor wake. The results indicate that more than 80\% of the turbulent kinetic energy is resolved throughout large parts of the wake, which is generally considered sufficient for scale-resolving simulations (\cite{pope2001turbulent}).

\subsection{Adjoint Investigation}
\label{subsec:adjoint_analysis}
The statistically converged flow field is subsequently passed to the adjoint solver. Three objective functions are considered, corresponding to rotor drag, rotor lift, and a combined drift objective representing equal contributions of drag and lift. According to Eqn.~\eqref{equ:adwall_bc}, the associated adjoint boundary conditions are obtained by prescribing $r_i=\delta_{i1}$, $r_i=\delta_{i2}$, and $r_i=[1,1]^\mathrm{T}/\sqrt{2}$, respectively. The steady adjoint calculations converge after approximately 5000 outer iterations.

The topological sensitivity field is obtained from the inner product of the statistically converged primal velocity field and its adjoint counterpart according to Eqn.~\eqref{equ:topo_derivative}. Following the conventional interpretation used in gradient-based topology optimization, the sensitivity field is multiplied by $-1$ such that positive values indicate regions where additional material is expected to improve the respective objective function.

\begin{figure}[!ht]
\centering
\subfigure[]{
\iftoggle{tikzExternal}{
\input{./tikz/sens_pos_drag_h1.tikz}}{
\includegraphics{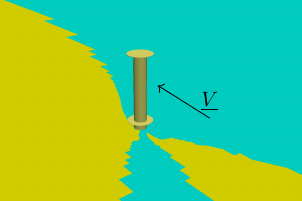}
}
}
\subfigure[]{
\iftoggle{tikzExternal}{
\input{./tikz/sens_pos_lift_h1.tikz}}{
\includegraphics{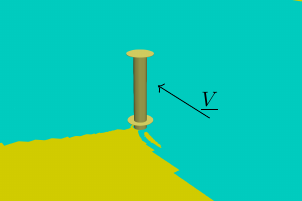}
}
}
\subfigure[]{
\iftoggle{tikzExternal}{
\input{./tikz/sens_pos_drift_h1.tikz}}{
\includegraphics{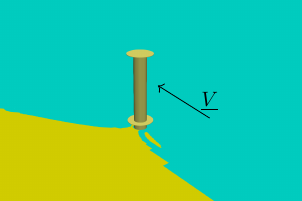}
}
}
\subfigure[]{
\iftoggle{tikzExternal}{
\input{./tikz/sens_pos_drag_h2.tikz}}{
\includegraphics{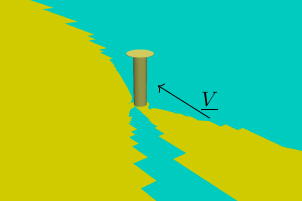}
}
}
\subfigure[]{
\iftoggle{tikzExternal}{
\input{./tikz/sens_pos_lift_h2.tikz}}{
\includegraphics{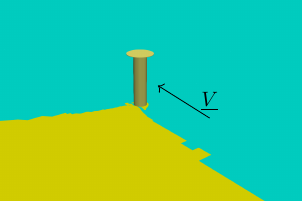}
}
}
\subfigure[]{
\iftoggle{tikzExternal}{
\input{./tikz/sens_pos_drift_h2.tikz}}{
\includegraphics{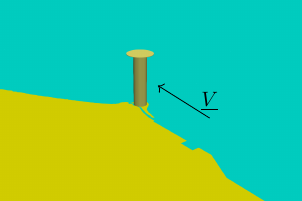}
}
}
\subfigure[]{
\iftoggle{tikzExternal}{
\input{./tikz/sens_pos_drag_h3.tikz}}{
\includegraphics{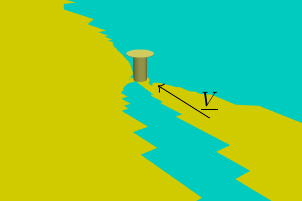}
}
}
\subfigure[]{
\iftoggle{tikzExternal}{
\input{./tikz/sens_pos_lift_h3.tikz}}{
\includegraphics{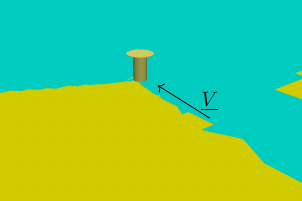}
}
}
\subfigure[]{
\iftoggle{tikzExternal}{
\input{./tikz/sens_pos_drift_h3.tikz}}{
\includegraphics{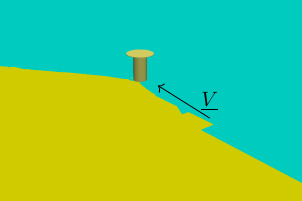}
}
}
\caption{Topological sensitivity fields for the drag, lift, and drift objectives (columns) evaluated on three horizontal planes at $x_3/H=[0,1,2]/3$ (rows). Positive (yellow) sensitivities indicate regions where additional material is expected to improve the respective objective function, whereas negative (turquoise) sensitivities indicate detrimental regions.}
\label{fig:des_topological_sensitivity}
\end{figure}

The resulting sensitivity distributions are shown in Fig.~\ref{fig:des_topological_sensitivity}, where positive (yellow) sensitivities indicate regions where additional material is expected to improve the respective objective function, whereas negative (turquoise) sensitivities indicate detrimental regions. Columns correspond to the drag, lift, and drift objectives, while rows represent three horizontal planes located at $x_3/H=[0,1,2]/3$. A comparison of the individual sections reveals only minor variations in the vertical direction, indicating that the overall sensitivity patterns are largely independent of height.

The lift and drift objectives exhibit very similar sensitivity distributions. In both cases, beneficial regions are predominantly located on the windward starboard side of the rotor, indicating that structures placed in these areas are likely to improve the corresponding objective function. The drift objective extends this favourable region slightly towards the areas identified by the drag objective. In contrast, the drag formulation predicts beneficial regions over large portions of the starboard side and upstream of the rotor. These observations provide the basis for the subsequent placement of container stacks used for validation of the proposed methodology.
\subsection{Validation}
\label{subsec:validation}

\begin{figure}[!ht]
\centering
\subfigure[]{
\includegraphics[scale=0.1]{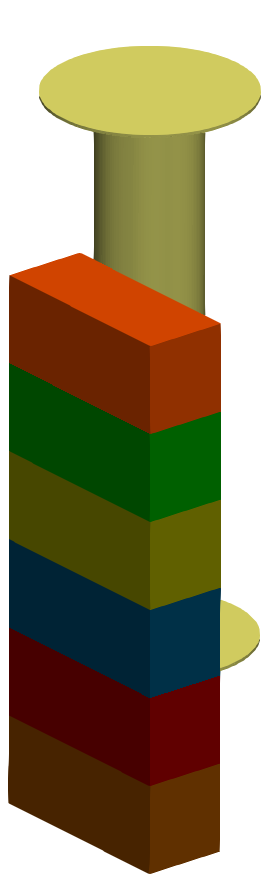}
}
\hspace{2.5cm}
\subfigure[]{
\includegraphics[scale=0.1]{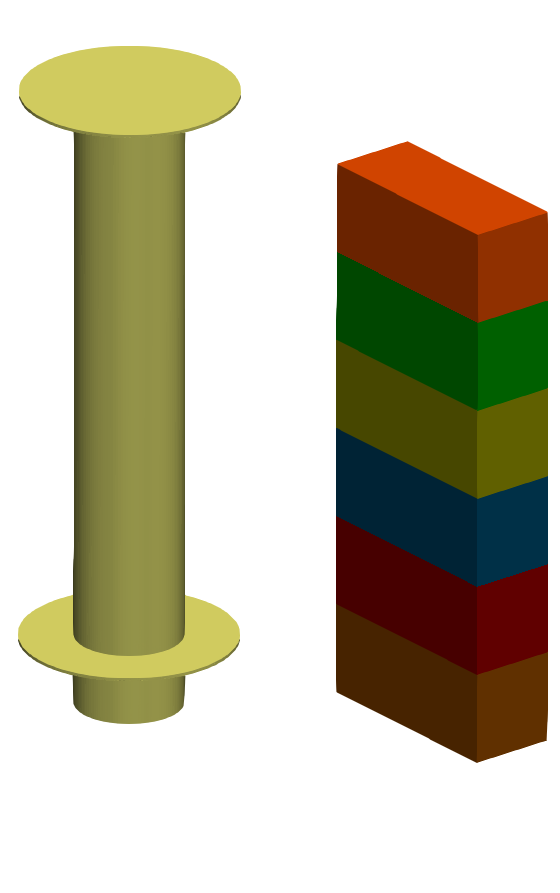}
}
\hspace{2.5cm}
\subfigure[]{
\includegraphics[scale=0.1]{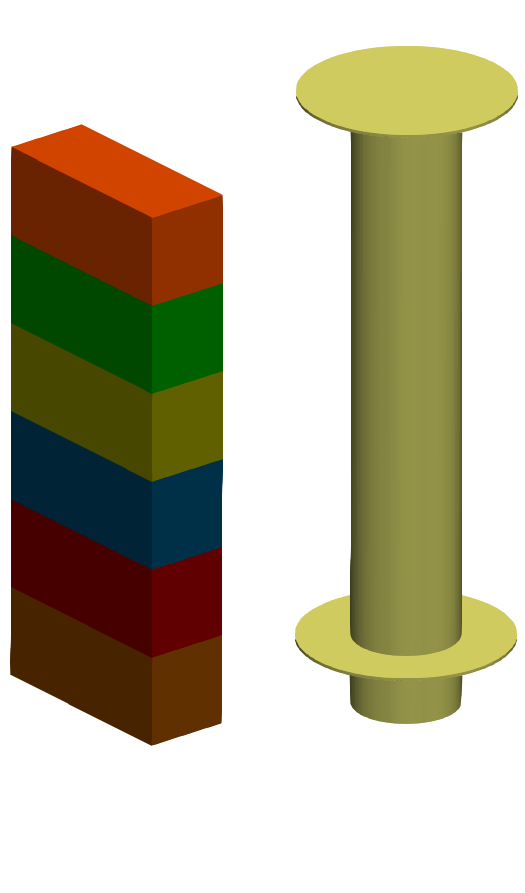}
}
\hspace{2.5cm}
\subfigure[]{
\includegraphics[scale=0.1]{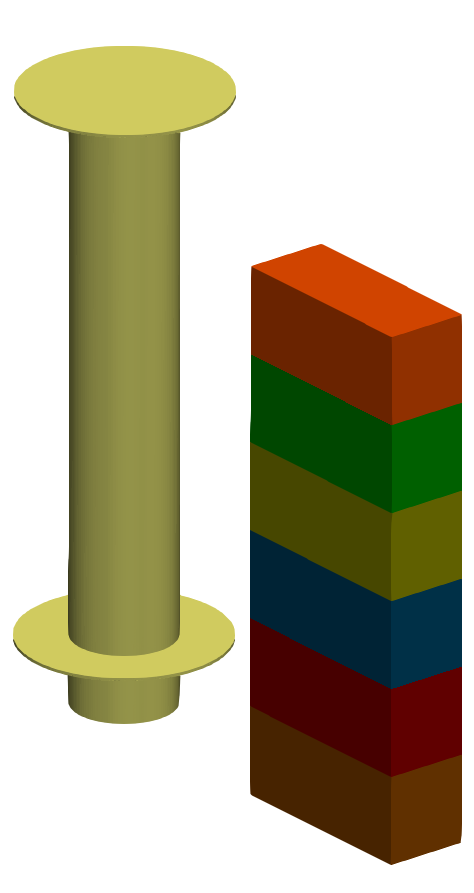}
}
\caption{Container-stack arrangements used for validation of the sensitivity predictions. Cases (a) and (b) assess the lift/drift objectives, whereas cases (c) and (d) evaluate the drag objective.}
\label{fig:container_arrangements}
\end{figure}

The predictive capability of the proposed sensitivity-based methodology is assessed by placing container stacks at locations identified as either beneficial or detrimental by the sensitivity maps. Four configurations are considered, each consisting of six stacked TEU20 containers with a total height of $h_\mathrm{cont.}/H=0.8636$.
Two arrangements are selected to validate the lift and drift objectives. The container stack is placed once within a region predicted to be beneficial on the starboard side of the rotor at $[x_1,x_2]/H=[1/5,-1/3]$ and once within a region predicted to be detrimental on the port side at $[x_1,x_2]/H=[1/5,1/3]$.
For the drag objective, two additional arrangements are investigated. In both cases, the container stack is positioned within regions predicted to improve the objective function, namely upstream of the rotor at $[x_1,x_2]/H=[1/3,0]$ and on the downstream starboard side at $[x_1,x_2]/H=[-1/3,-1/3]$. The resulting container–rotor configurations are shown in Fig.~\ref{fig:container_arrangements}.

All four validation cases are simulated using the numerical setup introduced for the reference configuration. Each simulation covers 200 equivalent rotor-flow passages, with statistical averaging starting after 100 passages. The additional container stacks increase the mesh size from approximately 3.8 million to 4.1 million control volumes.

\begin{figure}[!ht]
\centering
\iftoggle{tikzExternal}{
\tinyPicture

\begin{tikzpicture}
    \begin{axis}[
            ybar,
            bar width=10pt,
            width=0.4\textwidth,
            height=0.3\textwidth,
            legend style={at={(0.98,0.98)}, anchor=north east},
            ylabel style={text width=0.40\textwidth,align=center},
            legend columns=4,
            xtick=data,
            symbolic x coords={A,B,C,D},
            ymin=-60,
            ymax=60,
            ytick={-60, -40, -20, 0, 20, 40, 60},
            xlabel={arrangement},
            ylabel={$(J - J_\mathrm{ref})/J_\mathrm{ref} \cdot 100$ [\%]},
            xticklabels={(a), 
            		 (b),
            		 (c),
            		 (d),
            		},
            xticklabel style={rotate=00, anchor=north},
        ]
        
        \addplot [style={color1,fill=color1}] 	coordinates 	{ (A, 58.185) (B, 1.4612e+00) (C, -3.9807e+01) (D, -3.6121e+01) };
	\addplot [style={color2,fill=color2}] 	coordinates 	{ (A, 2.0078e+01) (B, -3.6660e+00) (C, -7.6955e+00) (D, -3.0076e+01) };

        \addlegendentry{drag};
        \addlegendentry{lift};
        
    \end{axis}
\end{tikzpicture}}{
\includegraphics{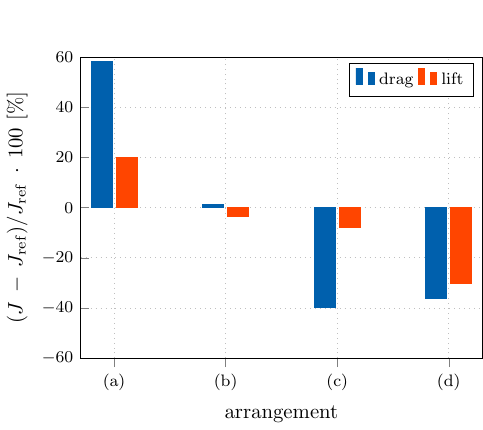}
}
\caption{Relative changes in drag and lift compared with the isolated rotor configuration for the four validation arrangements shown in Fig.~\ref{fig:container_arrangements}.}
\label{fig:final_force_results}
\end{figure}

The resulting force variations are summarized in Fig.~\ref{fig:final_force_results} relative to the reference configuration according to $(J-J_\mathrm{ref})/J_\mathrm{ref} \cdot 100\%$. Relative changes in drag and lift are shown separately for all four container arrangements considered.

The sensitivity-based predictions are confirmed by the subsequent flow simulations. For the lift/drift validation cases, the starboard arrangement (a), which was located within a beneficial sensitivity region, increases the lift coefficient, whereas the port-side arrangement (b), positioned within a detrimental region, reduces the lift coefficient. The predicted trend is therefore reproduced correctly in both cases.

The drag-oriented arrangements likewise confirm the sensitivity predictions. Both downstream-starboard (c) and upstream (d) container placements lead to a substantial reduction of the drag coefficient compared with the isolated rotor configuration. While both arrangements achieve comparable drag reductions, the downstream-starboard position results in a smaller penalty in lift.

Overall, the results demonstrate that the proposed sensitivity maps provide meaningful guidance for identifying favourable and unfavourable locations of surrounding structures. Despite the deliberately simplified interpretation of the sensitivity field, the predicted trends are consistently recovered by the corresponding flow simulations.

\section{Conclusion and Outlook}
\label{sec:conclusion_outlook}
A topology-optimization-inspired numerical sensitivity-analysis framework for identifying beneficial and detrimental structure locations in the vicinity of Flettner rotors has been presented. The methodology is based on adjoint-derived topological sensitivities obtained from a virtual porosity formulation. Rather than performing an actual topology optimization, the sensitivity field is interpreted qualitatively to identify regions where additional material is expected to improve or deteriorate a given aerodynamic objective.

The proposed approach has been demonstrated for a full-scale Flettner rotor operating under atmospheric wind conditions. Sensitivity maps were generated for drag, lift, and combined drift objectives and subsequently interpreted to derive candidate locations for container-stack placement. The resulting validation simulations confirmed the predicted trends, indicating that the proposed methodology provides meaningful guidance for the arrangement of surrounding structures in the vicinity of Flettner rotors.

From a practical perspective, the method may support early-stage ship design by rapidly identifying favourable and unfavourable regions for deck cargo, superstructures, and other aerodynamic appendages. Since the approach relies only on the sign of the sensitivity field, no optimization loop or geometric parameterization is required.

Future work should address more realistic operating conditions, including varying wind speeds and apparent wind angles. Furthermore, the methodology could be extended towards actual topology optimization procedures or applied to related maritime design problems involving aerodynamic and hydrodynamic flow-control devices.

\section{Declaration of Competing Interest}
The author declares that he has no known competing financial interests or personal relationships that could have appeared to influence the work reported in this paper.

\section{Acknowledgments}
The current work is part of the “Propulsion Optimization of Ships and Appendages” (Grant No. 03SX599C) and "Development of a Comprehensive Methodology for the Integration of Flettner Rotors on Different Ship Types" (Grant No. 03SX581G) research projects funded by the German Federal Ministry for Economics and Climate Action. The author gratefully acknowledges this support.

\section{Acknowledgment of AI Assistance}
OpenAI ChatGPT (GPT-5.5) was used to improve the wording and readability of parts of the manuscript. All scientific concepts, methodology, numerical simulations, analyses, and conclusions were developed and verified exclusively by the authors.

\section{Declaration of Competing Interest}
The authors declare that they have no known competing financial interests or personal relationships that could have appeared to influence the work reported in this paper.

\section{Data Availability Statement}
The data generated during the current study are available from the corresponding author upon reasonable request.



\end{document}